\documentclass[preprint,showpacs,preprintnumbers,amsmath,amssymb]{revtex4}

\usepackage{graphicx}

\begin{document}

\title{The Icosahedral Symmetry Antiferromagnetic Heisenberg Model}

\author{N.P. Konstantinidis}
\affiliation{Ames Laboratory and Department of Physics and Astronomy, Iowa
             State University, Ames, Iowa 50011}

\date{\today}

\begin{abstract}
The antiferromagnetic Heisenberg model on icosahedral symmetry $I_{h}$
fullerene clusters exhibits unconventional magnetic properties, despite the
lack of anisotropic interactions. At the classical level, and for number of
sites $n \leq 720$, the magnetization has two discontinuities in an external
magnetic field, except from the dodecahedron where it has three, emphasizing
the role of frustration introduced by the pentagons in the unusual magnetic
properties. For spin magnitudes $s_{i}=\frac{1}{2}$ there is a discontinuity
of quantum character close to saturation for $n \leq 80$. This common magnetic
behavior indicates that it is a generic feature of $I_{h}$ fullerene clusters,
irrespectively of $n$.
\end{abstract}


\maketitle

The antiferromagnetic Heisenberg model (AHM) is a prototype for strongly
correlated electronic behavior, and the combination of low dimensionality,
quantum
fluctuations and frustration produces unconventional magnetic behavior
\cite{Misguich03}. This includes non-magnetic excitations in the low-energy
spectrum, non-trivial dependence of the specific heat and the susceptibility
on temperature, and magnetization plateaux and discontinuities in a magnetic
field \cite{Waldtmann98,NPK05-2,Richter04,Honecker04}. Here the model is
investigated for spins sitting on vertices of clusters of the fullerene type,
with icosahedral spatial symmetry $I_{h}$ \cite{Altmann94}. Fullerene
molecules have been found to superconduct when doped with alkali metals
\cite{Hebard91,Holczer91}. An electronic mechanism for superconductivity was
suggested based on perturbation theory calculations on the one-band Hubbard
model on doped $C_{60}$, the fullerene with sixty carbon atoms and $I_{h}$
point group symmetry, which geometrically corresponds to the truncated
icosahedron \cite{Chakravarty91}. Diagonalization of the Hubbard model is
prohibitive due to limitations imposed by the dimensionality of the Hilbert
space of this molecule. As a first step, the model is considered on its
strong on-site repulsion limit at half-filling, the AHM, on clusters of
$I_{h}$ symmetry with number of sites $n$ up to $720$. We look for
correlations between the magnetic properties and spatial symmetry at the
classical and quantum level. The presence of such correlations could open
the possibility of studying smaller clusters to gain insight on larger ones
of the same symmetry, which are intractable with present day computational
means. This approach could as well be used for the Hubbard model to
investigate superconducting correlations.

Fullerene molecules are three-fold coordinated and consist of $12$ pentagons
and $\frac{n}{2}-10$ hexagons. For $I_{h}$ symmetry
the number of sites is given by $n_{1}=20 i^{2}$ (to be called type I
molecules) or $n_{2}=60 i^{2}$ (type II), with $i$ an integer
\cite{Fujita92,Yoshida97}. These clusters belong to the class of Goldberg
polyhedra. The smallest is the dodecahedron with $n=20$ and no hexagons, and
the largest considered here has $n=720$. Frustration is introduced by the
pentagons, with each surrounded only by hexagons for $n>20$. The dodecahedron
is the only cluster where pentagons are neighboring each other, hence
frustration is maximal. As the number of hexagons increases with the number
of sites, frustration on the average decreases. It has already been shown
that there are strong similarities in the low energy spectra, the specific
heat and the magnetic susceptibility of the dodecahedron and the icosahedron,
a five-fold coordinated cluster with the same spatial symmetry and $12$ sites
consisting only of triangles \cite{NPK05-1}. Here we extend the investigation
to the larger molecules and show that spatial symmetry determines to a large
extent the behavior in a magnetic field. We conclude that significant insight
can be gained on more complicated fermionic models on large $I_{h}$ fullerene
molecules from their solution on smaller molecules of the same symmetry.

The AHM is isotropic in spin space, and the total magnetization is usually a
smooth function of an external magnetic field at the classical level for
unfrustrated systems. Coffey and Trugman showed that it has a discontinuity
for the dodecahedron and the truncated icosahedron \cite{Coffey92}. It was
also shown that for individual spin magnitudes $s_{i}=\frac{1}{2}$ and $1$
the magnetization curve is discontinuous and the total spin changes by
$\Delta S=4 s_{i}$ for the dodecahedron, twice as much as the change between
adjacent $S$ sectors, with a particular sector never including the ground
state in the field \cite{NPK05-1}. The icosahedron also has a magnetization
jump for classical spins and for higher $s_{i}$ \cite{Schroder05}. The
icosidodecahedron is another molecule with $I_{h}$ symmetry, four-fold
coordinated and consisting of triangles and pentagons, with a similar
property at high magnetic fields for the lowest $s_{i}$ \cite{Schmidt05}.
Here the magnetic response of the AHM is calculated at the classical limit,
$s_{i} \to \infty$, and the full quantum limit, $s_{i}=\frac{1}{2}$. We find
that the response to an external magnetic field is discontinuous for all the
$I_{h}$ fullerene clusters, showing the correlation between magnetic behavior
and spatial symmetry. For classical spins, there are two discontinuities, one
at relatively small magnetic fields and the second at high fields close to
saturation. For the dodecahedron, another discontinuity precedes the low-field
one, bringing the total number to $3$, a rather uncommon feature in the
absence of anisotropic magnetic interactions. At the opposite limit,
$s_{i}=\frac{1}{2}$, there is a jump with $\Delta S=2$ for the three smallest
clusters ($n \leq 80$), where the lowest state of the sector with five
flipped spins from saturation and $S=\frac{n-5}{2}$ is never the ground state
in a field. The mechanism of the jump is the same in all cases. For higher
$n$'s memory requirements prohibit the calculation of the magnetization curve
down to the appropriate fields. These common properties of the $I_{h}$
fullerenes point out the importance of symmetry and frustration for the
determination of magnetic behavior.

The antiferromagnetic Heisenberg Hamiltonian for spins $\vec{s}_{i}$ on the
vertices $i$ of the clusters is
\begin{equation}
H = J \sum_{<i,j>} \textrm{} \vec{s}_{i} \cdot \vec{s}_{j} - h S^{z}
\label{eqn:1}
\end{equation}
where $<>$ denotes nearest neighbors, and $J$ is positive and is set equal
to $1$, defining the unit of energy. $h$ is the strength of an external
magnetic field in units of energy and $S^{z}$ the projection of the total
spin along the field direction $z$.

\begin{figure}
\includegraphics[width=5in,height=3.1in]{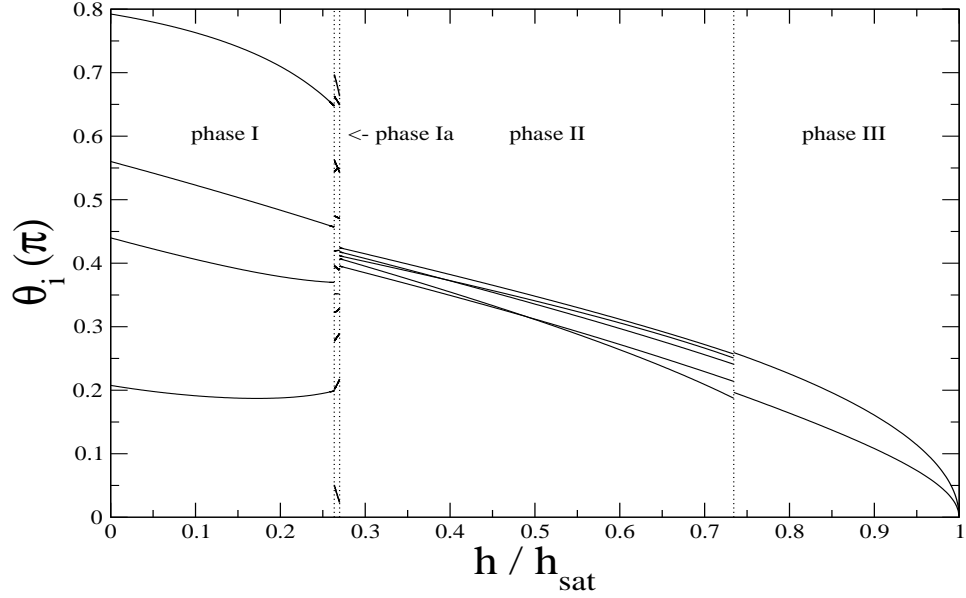}
\caption{Unique polar angles $\theta_{i}$ in units of $\pi$ in the ground state of the classical AHM on the dodecahedron, as a function of the magnetic field over its saturated value. The four phases are divided by vertical lines, which correspond to the transition fields given in table \ref{table:1}.}
\label{fig:1}
\end{figure}

For classical spins, $|\vec{s}_{i}|$ is taken equal to $1$, and the
Hamiltonian is a function of the spin polar and azimuthal angles. The
critical fields for the magnetization discontinuities are listed in table
\ref{table:1}. In the absence of a field the spins are non-coplanar in the
dodecahedron ground state \cite{Coffey92}. With increasing field, they turn
gradually to its direction, and are distributed around the $z$ axis at $4$
different polar angles in groups of $5$, as shown in figure \ref{fig:1}.
This phase is called I. Above the first critical field, the spins share a
common polar angle in pairs and the azimuthal angles of these pairs add up
to the same field-dependent value $c$, except from two pairs which have
different polar angles but the same azimuthal angle, equal to $\frac{c}{2}$
and $\frac{c}{2} + \pi$ respectively. This phase (Ia) is unique to the
dodecahedron and appears for a very small window of the magnetic field. In
the following phase (II), all the spins have now negative magnetic energy,
with polar angles not differing very much from each other that assume $5$
different values, each corresponding to $4$ spins (figure \ref{fig:1}). For
higher fields there is a third transition to a more symmetric phase III
around the field, where the spins form only $2$ polar angles with the $z$
axis in groups of $10$ until the magnetization saturates. The magnetization
curves of the other clusters differ only in the lack of phase Ia, and they
go directly from phase I to phase II at the first transition. All the
magnetization jumps are characterized by a discontinuous derivative of the
energy with respect to the field.  The number of different polar angles can
be expressed in terms of the integer $i$ that gives the number of sites
$n_{1}$ and $n_{2}$. For phase I it is $4i$, each corresponding to a group
of $5$ spins, plus $2i(i-1)$, each representing $10$ spins for clusters of
type I. For type II there are $6i^{2}$ different polar angles, each
corresponding to $10$ spins. For phase II there are $\frac{n}{4}$ different
polar angles each including $4$ spins for both types. For phase III the
corresponding numbers are $2i$, each having $10$ spins, plus $i(i-1)$, each
having $20$ spins for type I. For type II there are $2i$ different polar
angle values each corresponding to $10$ spins, plus $(3i-1)i$ values each
representing $20$ spins.

Table \ref{table:1} shows that as frustration decreases with $n$, phase II,
the least symmetric around the $z$ axis, becomes the ground state for a
wider range of fields. In contrast, phase I is suppressed to lower fields,
and phase III appears just before saturation. The change in the
magnetization $\Delta M$ over the saturation magnetization $n$ is
decreasing with $n$ for both transitions. These results indicate that the
discontinuities will occur for any value of $n$, albeit closer to $h=0$
and $h=h_{sat}$ and with smaller $\frac{\Delta M}{n}$ as $n$ increases.
The role of the pentagons, which introduce frustration in the system, is
not diminished even when the hexagons strongly prevail in number. It is
also noted that hysteresis curves were calculated by slowly increasing the
field from zero to saturation and then switching it off to zero in the
same manner. In both cases the magnetization curve is the same and no
hysteresis is observed. All the clusters are in phase I for low fields,
and in phase III for higher fields, with a transition having
discontinuous magnetic susceptibility between the two phases.

\begin{figure}
\includegraphics[width=5in,height=3.1in]{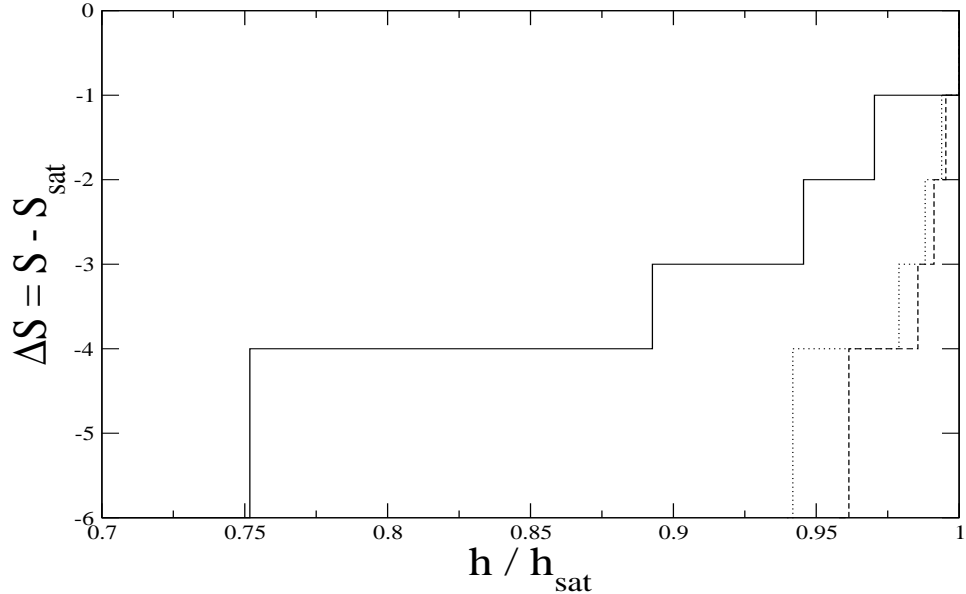}
\caption{Difference $\Delta S$ between the total spin in the ground state of the $s_{i}=\frac{1}{2}$ AHM, $S$,
and its saturated value $S_{sat}=\frac{n}{2}$, with $n$ the number of sites, as
a function of the magnetic field over its saturation value $\frac{h}{h_{sat}}$.
Solid line: $n=20$, dotted line: $n=60$, dashed line: $n=80$. The
discontinuities occur at $\frac{h_{c}}{h_{sat}} = 0.75177$, $0.94179$ and
$0.96142$ respectively.}
\label{fig:2}
\end{figure}

In the extreme quantum case $s_{i}=\frac{1}{2}$, the magnetization curve
typically follows a step-like structure with $\Delta S=1$ between adjacent
$S$ sectors. However, frustration can lead to magnetization
discontinuities. It has been found for the dodecahedron that the spin
sector $S=\frac{n-5}{2}$ with five flipped spins from saturation never
includes the ground state in a field, resulting in a step $\Delta S=2$
\cite{NPK05-1}. For $s_{i}=1$ a similar discontinuity was found, along
with a second one for lower $S$. The calculation is here extended to the
truncated icosahedron and the $n=80$ cluster, where memory requirements
permit Lanczos diagonalization for at least $S^{z}=\frac{n-6}{2}$.
Similarly to the dodecahedron, there is a magnetization discontinuity for
both clusters with a step $\Delta S=2$ between sectors on the sides of
$S=\frac{n-5}{2}$ (figure \ref{fig:2}). The discontinuities are
accompanied by magnetization plateaux. The mechanism of the jump is the
same in all three cases. The ground state below and above the transition
is non-degenerate. It switches from the $A_{g}$ to the $A_{u}$
one-dimensional irreducible representation of the $I_{h}$ symmetry group
as the field increases, changing its spatial symmetry from symmetric to
antisymmetric \cite{Altmann94}. For the dodecahedron, the magnetization
curve has also been calculated for higher quantum numbers up to
$s_{i}=\frac{5}{2}$ close to saturation, and there are no discontinuities
related to the sector with five flipped spins for $s_{i}>1$. The same is
true for the truncated icosahedron when $s_{i}=1$ or $\frac{3}{2}$. In
these cases there are only magnetization plateaux. This indicates that the
jump is not related to the classical limit, but rather is a purely quantum
effect. Similarly to the classical case, the discontinuities appear closer
and closer to the saturation field with increasing $n$.

The common magnetic properties of the AHM for all the $I_{h}$ fullerene
clusters investigated for $s_{i} \to \infty$ and $s_{i}=\frac{1}{2}$
suggest that they are shared by all the clusters of this class,
independently of $n$. Frustration, spatial symmetry and the presence of
pentagons result in magnetization discontinuities which are uncommon for
a model lacking magnetic anisotropy. Combined with the similarities in
the low energy spectra and thermodynamic properties of $I_{h}$ clusters
found in \cite{NPK05-1}, the results presented here show that predictions
for the behavior of fermionic models on $I_{h}$ clusters can be made by
studying smaller clusters of the same symmetry. In particular, comparison
of the energy of neutral $C_{60}$ plus two electrons to the total energy
of two separate molecules of neutral $C_{60}$ where one electron has been
added to each, shows if there is an effective attractive interaction
between the two electrons, favoring superconductivity \cite{Lin05}. This
calculation is much more tractable in the Hilbert space of the
dodecahedron, than the one of the truncated icosahedron.

In summary, the magnetization of the AHM for fullerene clusters of spatial
symmetry $I_{h}$ has been shown to exhibit discontinuities in a field
ranging up to $3$ at the classical level. The results indicate that the
discontinuities are a feature of any $I_{h}$ fullerene cluster, even
though phase II is strongly predominant with increasing $n$. For
$s_{i}=\frac{1}{2}$, there is also a jump for higher fields for
$n \leq 80$, which is of purely quantum character. Again it is anticipated
that this is a generic feature of the AHM on fullerene molecules with
$I_{h}$ symmetry. These effects are non-trivial in the absence of
anisotropic magnetic terms from the Hamiltonian. The common spatial
symmetry of the $I_{h}$ fullerenes leads to similar magnetic behavior. It
is of interest to examine correlations between spatial symmetry and
magnetic properties for other types of symmetry, as well as investigate
correlations between electronic behavior and spatial symmetry for more
complicated models with orbital degrees of freedom. The findings of this
article show that insight on the superconducting properties of the
truncated icosahedron can be gained from considering the significantly
smaller Hilbert space of the dodecahedron.

The author thanks D. Coffey, P. Herzig, M. Luban and L. Engelhardt for
discussions. Ames Laboratory is operated for the United States Department
of Energy by Iowa State University under contract No. W-7405-Eng-82.

\bibliography{papersix}

\begin{thebibliography}{16}
\expandafter\ifx\csname natexlab\endcsname\relax\def\natexlab#1{#1}\fi
\expandafter\ifx\csname bibnamefont\endcsname\relax
  \def\bibnamefont#1{#1}\fi
\expandafter\ifx\csname bibfnamefont\endcsname\relax
  \def\bibfnamefont#1{#1}\fi
\expandafter\ifx\csname citenamefont\endcsname\relax
  \def\citenamefont#1{#1}\fi
\expandafter\ifx\csname url\endcsname\relax
  \def\url#1{\texttt{#1}}\fi
\expandafter\ifx\csname urlprefix\endcsname\relax\def\urlprefix{URL }\fi
\providecommand{\bibinfo}[2]{#2}
\providecommand{\eprint}[2][]{\url{#2}}

\bibitem[{\citenamefont{Misguich and Lhuillier}()}]{Misguich03}
\bibinfo{author}{\bibfnamefont{G.}~\bibnamefont{Misguich}} \bibnamefont{and}
  \bibinfo{author}{\bibfnamefont{C.}~\bibnamefont{Lhuillier}}, \eprint{in {\it
  Frustrated Spin Systems}, edited by H.T. Diep (World Scientific, 2003).}

\bibitem[{\citenamefont{Waldtmann et~al.}(1998)\citenamefont{Waldtmann, Everts,
  Bernu, Lhuillier, Sindzingre, Lecheminant, and Pierre}}]{Waldtmann98}
\bibinfo{author}{\bibfnamefont{C.}~\bibnamefont{Waldtmann}},
  \bibinfo{author}{\bibfnamefont{H.-U.} \bibnamefont{Everts}},
  \bibinfo{author}{\bibfnamefont{B.}~\bibnamefont{Bernu}},
  \bibinfo{author}{\bibfnamefont{C.}~\bibnamefont{Lhuillier}},
  \bibinfo{author}{\bibfnamefont{P.}~\bibnamefont{Sindzingre}},
  \bibinfo{author}{\bibfnamefont{P.}~\bibnamefont{Lecheminant}},
  \bibnamefont{and} \bibinfo{author}{\bibfnamefont{L.}~\bibnamefont{Pierre}},
  \bibinfo{journal}{Eur. Phys. J. C} \textbf{\bibinfo{volume}{2}},
  \bibinfo{pages}{501} (\bibinfo{year}{1998}).

\bibitem[{\citenamefont{Konstantinidis}()}]{NPK05-2}
\bibinfo{author}{\bibfnamefont{N.~P.} \bibnamefont{Konstantinidis}},
  \eprint{cond-mat/0503659}.

\bibitem[{\citenamefont{Richter et~al.}(2004)\citenamefont{Richter,
  Schulenburg, Honecker, Schnack, and Schmidt}}]{Richter04}
\bibinfo{author}{\bibfnamefont{J.}~\bibnamefont{Richter}},
  \bibinfo{author}{\bibfnamefont{J.}~\bibnamefont{Schulenburg}},
  \bibinfo{author}{\bibfnamefont{A.}~\bibnamefont{Honecker}},
  \bibinfo{author}{\bibfnamefont{J.}~\bibnamefont{Schnack}}, \bibnamefont{and}
  \bibinfo{author}{\bibfnamefont{H.-J.} \bibnamefont{Schmidt}},
  \bibinfo{journal}{J. Phys. Cond. Matt.} \textbf{\bibinfo{volume}{16}},
  \bibinfo{pages}{779} (\bibinfo{year}{2004}).

\bibitem[{\citenamefont{Honecker et~al.}(2004)\citenamefont{Honecker,
  Schulenburg, and Richter}}]{Honecker04}
\bibinfo{author}{\bibfnamefont{A.}~\bibnamefont{Honecker}},
  \bibinfo{author}{\bibfnamefont{J.}~\bibnamefont{Schulenburg}},
  \bibnamefont{and} \bibinfo{author}{\bibfnamefont{J.}~\bibnamefont{Richter}},
  \bibinfo{journal}{J. Phys. Cond. Matt.} \textbf{\bibinfo{volume}{16}},
  \bibinfo{pages}{749} (\bibinfo{year}{2004}).

\bibitem[{\citenamefont{Altmann and Herzig}()}]{Altmann94}
\bibinfo{author}{\bibfnamefont{S.~L.} \bibnamefont{Altmann}} \bibnamefont{and}
  \bibinfo{author}{\bibfnamefont{P.}~\bibnamefont{Herzig}}, \eprint{{\it
  Point-Group Theory Tables} (Oxford University Press, 1994)}.

\bibitem[{\citenamefont{Hebard et~al.}(1991)\citenamefont{Hebard, Roseeinsky,
  Haddon, Murphy, Glarum, Palstra, Ramirez, and Kortan}}]{Hebard91}
\bibinfo{author}{\bibfnamefont{A.~F.} \bibnamefont{Hebard}},
  \bibinfo{author}{\bibfnamefont{M.~J.} \bibnamefont{Roseeinsky}},
  \bibinfo{author}{\bibfnamefont{R.~C.} \bibnamefont{Haddon}},
  \bibinfo{author}{\bibfnamefont{D.~W.} \bibnamefont{Murphy}},
  \bibinfo{author}{\bibfnamefont{S.~H.} \bibnamefont{Glarum}},
  \bibinfo{author}{\bibfnamefont{T.~T.~M.} \bibnamefont{Palstra}},
  \bibinfo{author}{\bibfnamefont{A.~P.} \bibnamefont{Ramirez}},
  \bibnamefont{and} \bibinfo{author}{\bibfnamefont{A.~R.}
  \bibnamefont{Kortan}}, \bibinfo{journal}{Nature (London)}
  \textbf{\bibinfo{volume}{350}}, \bibinfo{pages}{600} (\bibinfo{year}{1991}).

\bibitem[{\citenamefont{Holczer et~al.}(1991)\citenamefont{Holczer, Klein,
  Huang, Kaner, Fu, Whetten, and Diederich}}]{Holczer91}
\bibinfo{author}{\bibfnamefont{K.}~\bibnamefont{Holczer}},
  \bibinfo{author}{\bibfnamefont{O.}~\bibnamefont{Klein}},
  \bibinfo{author}{\bibfnamefont{S.-M.} \bibnamefont{Huang}},
  \bibinfo{author}{\bibfnamefont{R.~B.} \bibnamefont{Kaner}},
  \bibinfo{author}{\bibfnamefont{K.-J.} \bibnamefont{Fu}},
  \bibinfo{author}{\bibfnamefont{R.~L.} \bibnamefont{Whetten}},
  \bibnamefont{and}
  \bibinfo{author}{\bibfnamefont{F.}~\bibnamefont{Diederich}},
  \bibinfo{journal}{Science} \textbf{\bibinfo{volume}{252}},
  \bibinfo{pages}{1154} (\bibinfo{year}{1991}).

\bibitem[{\citenamefont{Chakravarty et~al.}(1991)\citenamefont{Chakravarty,
  Gelfand, and Kivelson}}]{Chakravarty91}
\bibinfo{author}{\bibfnamefont{S.}~\bibnamefont{Chakravarty}},
  \bibinfo{author}{\bibfnamefont{M.}~\bibnamefont{Gelfand}}, \bibnamefont{and}
  \bibinfo{author}{\bibfnamefont{S.}~\bibnamefont{Kivelson}},
  \bibinfo{journal}{Science} \textbf{\bibinfo{volume}{254}},
  \bibinfo{pages}{970} (\bibinfo{year}{1991}).

\bibitem[{\citenamefont{Fujita et~al.}(1992)\citenamefont{Fujita, Saito,
  Dresselhaus, and Dresselhaus}}]{Fujita92}
\bibinfo{author}{\bibfnamefont{M.}~\bibnamefont{Fujita}},
  \bibinfo{author}{\bibfnamefont{R.}~\bibnamefont{Saito}},
  \bibinfo{author}{\bibfnamefont{G.}~\bibnamefont{Dresselhaus}},
  \bibnamefont{and} \bibinfo{author}{\bibfnamefont{M.~S.}
  \bibnamefont{Dresselhaus}}, \bibinfo{journal}{Phys. Rev. B (R)}
  \textbf{\bibinfo{volume}{45}}, \bibinfo{pages}{13834} (\bibinfo{year}{1992}).

\bibitem[{\citenamefont{Yoshida}()}]{Yoshida97}
\bibinfo{author}{\bibfnamefont{M.}~\bibnamefont{Yoshida}},
  \eprint{http://cochem2.tutkie.tut.ac.jp:8000/Fuller/}.

\bibitem[{\citenamefont{Konstantinidis}(2005)}]{NPK05-1}
\bibinfo{author}{\bibfnamefont{N.~P.} \bibnamefont{Konstantinidis}},
  \bibinfo{journal}{Phys. Rev. B} \textbf{\bibinfo{volume}{72}},
  \bibinfo{pages}{064453} (\bibinfo{year}{2005}).

\bibitem[{\citenamefont{Coffey and Trugman}(1992)}]{Coffey92}
\bibinfo{author}{\bibfnamefont{D.}~\bibnamefont{Coffey}} \bibnamefont{and}
  \bibinfo{author}{\bibfnamefont{S.~A.} \bibnamefont{Trugman}},
  \bibinfo{journal}{Phys. Rev. Lett.} \textbf{\bibinfo{volume}{69}},
  \bibinfo{pages}{176} (\bibinfo{year}{1992}).

\bibitem[{\citenamefont{Schr{\"o}der et~al.}(2005)\citenamefont{Schr{\"o}der,
  Schmidt, Schnack, and Luban}}]{Schroder05}
\bibinfo{author}{\bibfnamefont{C.}~\bibnamefont{Schr{\"o}der}},
  \bibinfo{author}{\bibfnamefont{H.-J.} \bibnamefont{Schmidt}},
  \bibinfo{author}{\bibfnamefont{J.}~\bibnamefont{Schnack}}, \bibnamefont{and}
  \bibinfo{author}{\bibfnamefont{M.}~\bibnamefont{Luban}},
  \bibinfo{journal}{Phys. Rev. Lett.} \textbf{\bibinfo{volume}{94}},
  \bibinfo{pages}{207203} (\bibinfo{year}{2005}).

\bibitem[{\citenamefont{Schmidt et~al.}(2005)\citenamefont{Schmidt, Richter,
  and Schnack}}]{Schmidt05}
\bibinfo{author}{\bibfnamefont{R.}~\bibnamefont{Schmidt}},
  \bibinfo{author}{\bibfnamefont{J.}~\bibnamefont{Richter}}, \bibnamefont{and}
  \bibinfo{author}{\bibfnamefont{J.}~\bibnamefont{Schnack}},
  \bibinfo{journal}{J. Magn. Magn. Mater.} \textbf{\bibinfo{volume}{295}},
  \bibinfo{pages}{164} (\bibinfo{year}{2005}).

\bibitem[{\citenamefont{Lin et~al.}(2005)\citenamefont{Lin, {\v S}makov,
  S{\o}rensen, Kallin, and Berlinsky}}]{Lin05}
\bibinfo{author}{\bibfnamefont{F.}~\bibnamefont{Lin}},
  \bibinfo{author}{\bibfnamefont{J.}~\bibnamefont{{\v S}makov}},
  \bibinfo{author}{\bibfnamefont{E.~S.} \bibnamefont{S{\o}rensen}},
  \bibinfo{author}{\bibfnamefont{C.}~\bibnamefont{Kallin}}, \bibnamefont{and}
  \bibinfo{author}{\bibfnamefont{A.~J.} \bibnamefont{Berlinsky}},
  \bibinfo{journal}{Phys. Rev. B} \textbf{\bibinfo{volume}{71}},
  \bibinfo{pages}{165436} (\bibinfo{year}{2005}).

\end{thebibliography}


\begin{table}[h]
\begin{center}
\caption{Magnetic fields over their saturation value $\frac{h_{c}}{h_{sat}}$
with a magnetization discontinuity for $n$ sites, and reduced magnetization
values on the sides of the discontinuity $\frac{M_{-}}{n}$ and
$\frac{M_{+}}{n}$. For the dodecahedron ($n=20$), the numbers in the left part
of the table correspond to the phase Ia to phase II transition. The
corresponding numbers for the phase I to phase Ia transition are $0.26350$,
$0.22411$ and $0.22660$.}
\hspace{18pt} phase $I -> II$ \hspace{35pt} phase $II -> III$

\begin{tabular}{c|c|c|c|c|c|c}
$n$ & $\frac{h_{c}}{h_{sat}}$ & $\frac{M_{-}}{n}$ & $\frac{M_{+}}{n}$ &
$\frac{h_{c}}{h_{sat}}$ & $\frac{M_{-}}{n}$ & $\frac{M_{+}}{n}$ \\
\hline
 20 & 0.26982 & 0.23688 & 0.27518 & 0.73428 & 0.74766 & 0.75079 \\
\hline
 60 & 0.14692 & 0.11723 & 0.14790 & 0.94165 & 0.94574 & 0.94651 \\
\hline
 80 & 0.12596 & 0.10003 & 0.12702 & 0.95134 & 0.95465 & 0.95527 \\
\hline
180 & 0.08275 & 0.06438 & 0.08325 & 0.98013 & 0.98208 & 0.98234 \\
\hline
240 & 0.07139 & 0.05525 & 0.07176 & 0.98541 & 0.98697 & 0.98716 \\
\hline
320 & 0.06161 & 0.04749 & 0.06190 & 0.98902 & 0.99027 & 0.99041 \\
\hline
500 & 0.04910 & 0.03765 & 0.04928 & 0.99301 & 0.99386 & 0.99396 \\
\hline
540 & 0.04722 & 0.03618 & 0.04738 & 0.99355 & 0.99434 & 0.99443 \\
\hline
720 & 0.04082 & 0.03120 & 0.04094 & 0.99516 & 0.99577 & 0.99584 \\
\end{tabular}
\label{table:1}
\end{center}
\end{table}

\newpage

\vspace{10pt}

\end{document}